\documentclass[aps,prd,amsfonts,amssymb, amsmath, showpacs, showkeys, a4paper, nofootinbib, superscriptaddress, 11pt, raggedbottom]{revtex4-2}
\usepackage{amsmath,amssymb}
\usepackage[colorlinks=true,urlcolor=blue,citecolor=blue,linkcolor=blue]{hyperref}
\usepackage{graphicx}
\usepackage{braket}
\usepackage{subcaption}
\usepackage[T1]{fontenc}

\usepackage{soul}
\usepackage{xcolor}

\begin{document}

\title{Quantum-enhanced screened dark energy detection}

\author{Daniel Hartley}
\email{danielhartley0@yahoo.com.au}
\affiliation{Faculty of Physics, University of Vienna, Boltzmanngasse 5, 1090 Wien, Austria}
\author{Christian K{\"a}ding}
\email{christian.kaeding@tuwien.ac.at}
\affiliation{Technische Universit\"at Wien, Atominstitut, Stadionallee 2, 1020 Vienna, Austria}
\affiliation{National Research University Higher School of Economics, 101000, Moscow, Russia}
\affiliation{School of Physics \& Astronomy, University of Nottingham, University Park, Nottingham NG7 2RD, United Kingdom}
\author{Richard Howl}
\email{richard.howl@rhul.ac.uk}
\affiliation{Department of Physics, Royal Holloway, University of London, Egham, Surrey, TW20 0EX, United Kingdom}
\affiliation{Quantum Group, Department of Computer Science, University of Oxford,
Wolfson Building, Parks Road, Oxford, OX1 3QD, United Kingdom}
\affiliation{QICI Quantum Information and Computation Initiative, Department of Computer Science,
The University of Hong Kong, Pokfulam Road, Hong Kong}
\affiliation{School of Mathematical Sciences, University of Nottingham, University Park, Nottingham NG7 2RD, United Kingdom}
\author{Ivette Fuentes}
\email{I.Fuentes-Guridi@soton.ac.uk}
\affiliation{School of Physics and Astronomy, University of Southampton, Southampton SO17 1BJ, United Kingdom}
\affiliation{Keble College, Oxford OX1 3PG, United Kingdom}
\affiliation{School of Mathematical Sciences, University of Nottingham, University Park, Nottingham NG7 2RD, United Kingdom}


\begin{abstract}
We propose an experiment based on a Bose-Einstein condensate interferometer for strongly constraining fifth-force models. Additional scalar fields from modified gravity or higher dimensional theories may account for dark energy and the accelerating expansion of the Universe. These theories have led to proposed screening mechanisms to fit within the tight experimental bounds on fifth-force searches. We show that our proposed experiment would greatly improve the existing constraints on these screening models by many orders of magnitude.
\end{abstract}

\maketitle


\section{Introduction}

General relativity (GR) has remained a tremendously successful theory, producing accurate physical predictions consistent with the barrage of experiments and observations conducted over the last century.
Despite this success, there are still many open problems within GR and apparent limitations of the theory itself.
Amongst modified theories of gravity aiming to address these problems, scalar-tensor theories (e.g.\,\,Brans-Dicke theory \cite{Brans1961}, see also \cite{Fujii2003}) are some of the most widely studied.
Modified theories of gravity like $f(R)$-gravity can additionally be shown to be equivalent to scalar-tensor theories, and higher dimensional theories (e.g.\,\,string theory) predict the existence of effective scalar field modes in 4-dimensional spacetime due to compactifications of the extra dimensions \cite{Wehus2002}.

Modifications of gravity gained even greater attention after the accelerated expansion of the Universe was discovered \cite{Perlmutter1998,Riess1998} and the puzzle of dark energy (DE) - the energy that supposedly drives this expansion - arose.
Consequently, there have been several proposed explanations for the nature of DE based on scalar-tensor theories (see e.g.\,\,\cite{Clifton2011,Joyce2014} for an overview of models).
Some of these models are predicted to cause a fifth force, which, at first glance, seems to be in contradiction with observations and experiments \cite{Dickey1994,Adelberger2003,Kapner2007}. While, consequently, some of these models have already been ruled out by observations \cite{Ishak2018}, those with a so-called screening mechanism \cite{Burrage2017} have features that suppress the effects of the additional scalar fields in regions of high matter density, such that they may contribute to dark energy while the coupling to matter as a fifth force still evades experimental constraints. What constitutes a high or low matter density is strongly dependent on the scalar field model parameters. As a rule of thumb, it is sensible to say that, for a given set of model parameters, a mass density is high/low if it leads to the considered scalar field model being screened/unscreened. Certainly, in order to avoid the abovementioned constraints, the average density of our Solar System must be assumed to be high.

Cold atom systems have proven to be invaluable tools in precision metrology. From practical applications such as ultra-high precision clocks \cite{Ludlow2015} to more fundamental experiments searching e.g.\,\,for deviation from the equivalence principle \cite{Schlippert2015,Overstreet2018,Becker2018}, the high degree of control and low internal noise afforded by cold atom systems makes them an ideal testing ground. Many scalar-tensor theories assume a conformal coupling between the metric tensor and the scalar field, and cold atom systems have been found to be well suited to studying these particular models in experiments (e.g.\,\,in atom interferometers \cite{Copeland2014,Jaffe2017}) and analogue gravity simulations \cite{Hartley2019}. In addition, there have even been proposals for testing the open quantum dynamics induced by such scalar fields in superposed cold atoms \cite{Kading1,Kading2}.

In this article, we propose using a guided Bose-Einstein condensate\footnote{We consider BECs as they admit the simplest theoretical analysis, but are still widely experimentally implemented. It is possible that thermal states or Fermi gases could also yield promising results, but we leave these investigations to future work.} (BEC) interferometer scheme to further constrain these conformally coupled screened scalar field models. Guided is used in this context to refer to atoms held in a trap for all or most of the interferometer scheme, rather than being in free fall. For this scheme, we consider a guided BEC interferometer as currently demonstrated in experiments. The main advantage of this scheme is a longer integration time: a trapped BEC can be held near a source object for much longer than atoms in a ballistic trajectory. We show that the constraints on the above screened scalar field models could be improved by many orders of magnitude.


\section{Scalar field models}
\label{sec:Models}

The models we consider here come from scalar-tensor theories of gravity \cite{Fujii2003}. As stated above, an additional scalar field $\varphi$ may be coupled to the metric tensor conformally in these theories, such that ordinary matter fields evolve according to the conformal metric
\begin{eqnarray}
\tilde{g}_{\mu\nu}&=&A^2\left(\varphi\right)g_{\mu\nu}
\end{eqnarray}
for some conformal factor $A^2\left(\varphi\right)$, where $g_{\mu\nu}$ is the normal GR metric. The equilibrium state of the $\varphi$ field is determined by minimising an effective potential \cite{Joyce2014,Burrage2017,Khoury2003}
\begin{eqnarray}\label{eq:EffPot}
V_\text{eff}\left(\varphi\right)&=&V\left(\varphi\right)+A\left(\varphi\right)\rho\,\,\,,
\end{eqnarray}
where $V\left(\varphi\right)$ is the self-interaction potential of the model and $\rho$ is the ordinary matter density.

We specifically consider two prominent examples of fifth force models with screening mechanism, namely the chameleon field \cite{Khoury20032,Khoury2003} and the symmetron field (first described in \cite{Dehnen1992, Gessner1992, Damour1994, Pietroni2005, Olive2008, Brax2010} and introduced with its current name in \cite{Hinterbichler2010,Hinterbichler2011}). These models have been investigated in atom interferometry experiments since, as is also the case with any other experiment performed in a vacuum chamber, a sufficiently thick wall of a vacuum chamber can shield the interior chameleon or symmetron scalar field from outside effects \cite{Copeland2014,Burrage2016}. More precisely, the scalar field is screened within the chamber walls, such that it creates a boundary separating the field outside and inside the chamber, which in turn leads to an evasion of any communication between both separated parts. As a consequence, the field inside the chamber is not affected by any mass densities outside the vacuum chamber. This allows the chamber's ultra-high vacuum to simulate the low density conditions of empty space resulting in long range (and thus measurable) chameleon or symmetron forces. Note that by a long-ranged force we mean one that falls off around a spherical source with $1/r^n$ for the radial component $r$ and some real number $n$ \cite{Nowakowski:2000zd}, and whose magnitude close to the source is comparable to or larger than gravity. In contrast, a short-ranged force falls off significantly faster than $1/r^n$ or is extremely weak even close to its source.


\subsection{Chameleons}

The chameleon field model is described by the conformal coupling \cite{Khoury2003}
\begin{eqnarray}\label{eq:ChamCoup}
A^2\left(\varphi\right)&=&\exp\left[\varphi/M_c\right]\,\,\,,
\end{eqnarray}
and the potential
\begin{eqnarray}\label{eq:ChamPot}
V\left(\varphi\right)&=&\Lambda^4\exp\left[\Lambda^{n}/\varphi^n\right].
\end{eqnarray}
The parameter $M_c$ determines the strength of the chameleon-matter coupling. This parameter is essentially unconstrained but is plausibly below the reduced Planck mass $M_{Pl}\approx 2.4\times10^{18}$ GeV/$c^2$. The self-interaction strength $\Lambda$ determines the contribution of the chameleon field to the energy density of the Universe, as the potential can be expanded as $V\approx\Lambda^4+\Lambda^{4+n}/\varphi^n$. This energy density can drive the accelerated expansion of the Universe observed today if $\Lambda=\Lambda_{DE}\approx 2.4$ meV. Finally, different choices of the parameter $n$ define different models, where $n\in\mathbb{Z}^+\cup\left\{x:-1<x<0\right\}\cup2\mathbb{Z}^-\backslash\left\{-2\right\}$ produces valid models with screening mechanisms. The two most commonly studied chameleon models are those where $n=1$ or $-4$ \cite{Burrage2017}.

\begin{figure}[htbp]
\begin{center}
\begin{subfigure}[b]{0.45\textwidth}
\includegraphics[width=\textwidth]{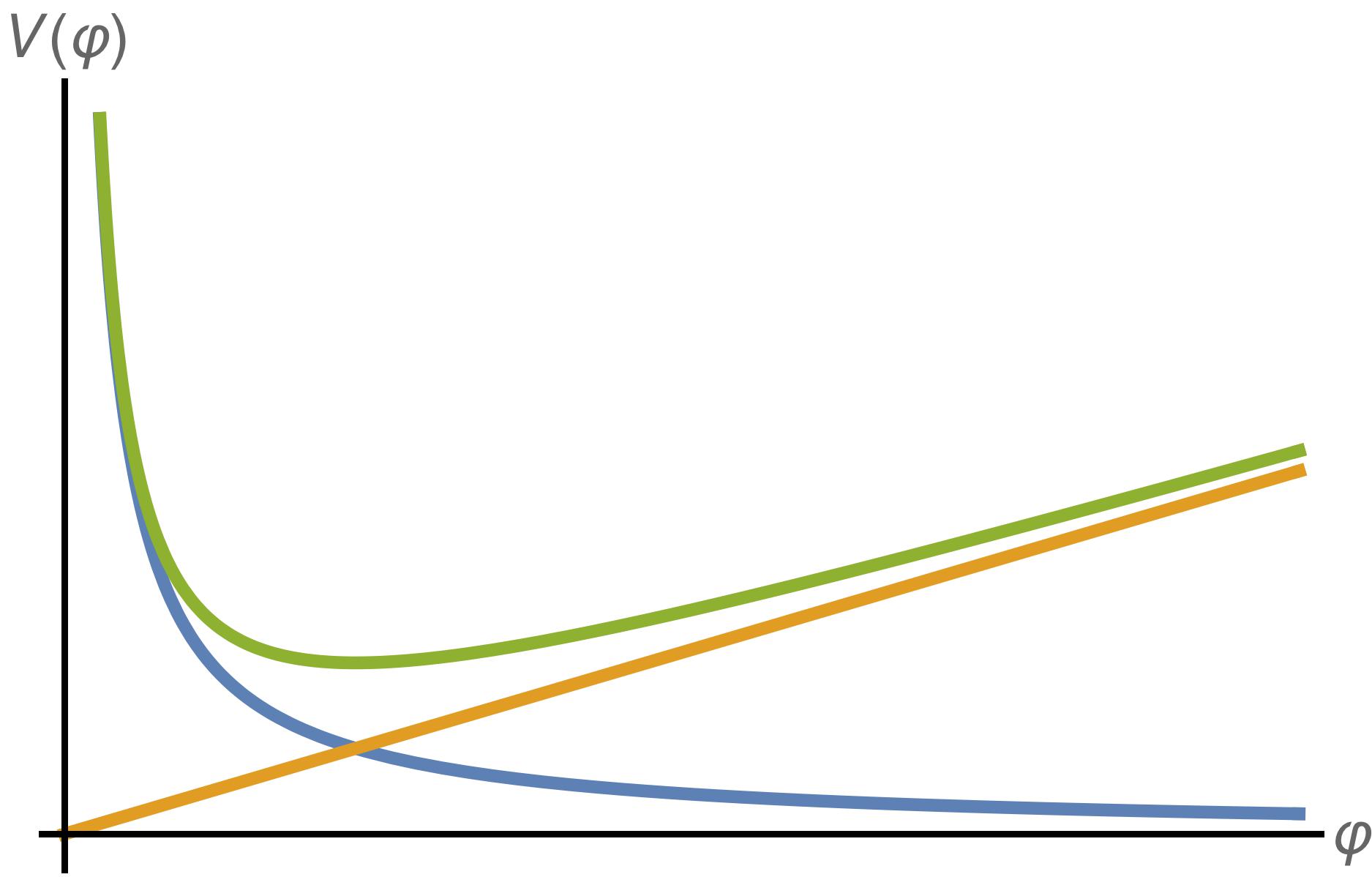}
\caption{}
\label{subFig:chameleonlow}
\end{subfigure}
\begin{subfigure}[b]{0.45\textwidth}
\includegraphics[width=\textwidth]{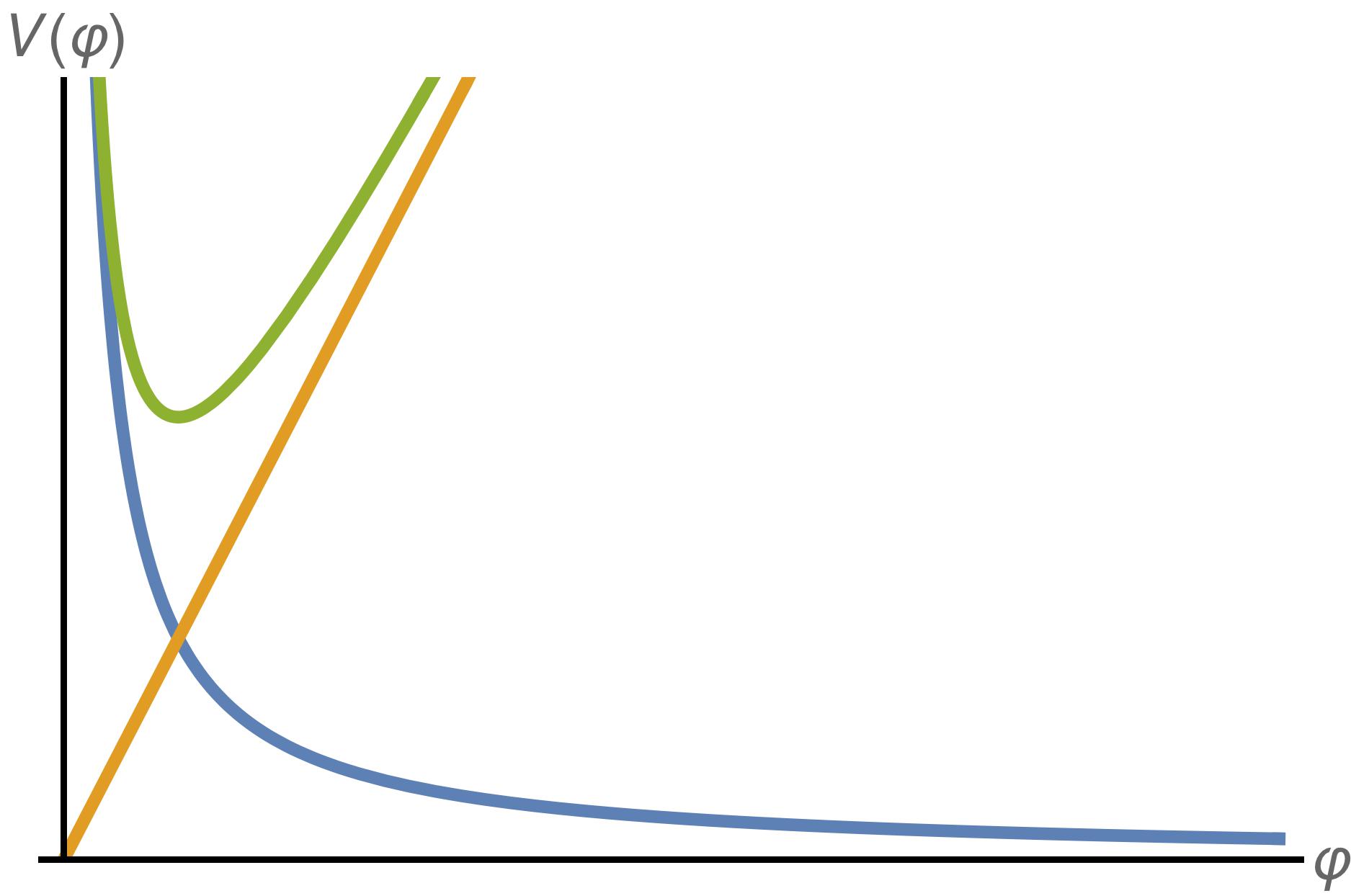}
\caption{}
\label{subFig:chameleonhigh}
\end{subfigure}
\caption{$n=1$ chameleon effective potential, see Eq.\,\,(\ref{eq:ChamEffPot}), for high (left) and low (right) ordinary matter densities plotted in green with its components: self-interaction (blue) and matter coupling (orange) }
\label{Fig:veff_chameleon}
\end{center}
\end{figure}

Following Eq.\,(\ref{eq:EffPot}), the conformal coupling in Eq.\,(\ref{eq:ChamCoup}) and the potential in Eq.\,(\ref{eq:ChamPot}) give rise to an effective chameleon potential, which is given to lowest order by
\begin{eqnarray}\label{eq:ChamEffPot}
V_\text{eff}\left(\varphi\right)&=&\frac{\Lambda^{4+n}}{\varphi^n}+\frac{\rho}{2M_c}\varphi+\mathcal{O}\left(\frac{\varphi^2}{M_c^2}\right)\,\,\,.
\end{eqnarray}
Fig.\,\ref{Fig:veff_chameleon} compares this effective potential (in green) in high and low density environments for the $n=1$ chameleon. In addition, it depicts its two components in blue and orange respectively for low (Fig.\,\ref{subFig:chameleonlow}) and high (Fig.\,\ref{subFig:chameleonhigh}) values of $\rho$. It is reasonable to ignore higher order terms in $\varphi/M_c$ in Eq.\,(\ref{eq:ChamEffPot}) as any fifth force effect measured on or near the Earth must be perturbative to be consistent with experimental observations.

The effective mass of the chameleon field in equilibrium is determined by the minimum of its effective potential, i.e. $m_c^2=\left|\partial^2 V_\text{eff}/\partial\varphi^2\right|_{\varphi=\varphi_\text{min}}$. The position of the effective potential minimum (and thus effective mass) depends on the ordinary matter density $\rho$. For example, the effective mass of excitations for the $n=1$ chameleon field is given by
\begin{eqnarray}
m_c^2&=&\left|\frac{\partial^2V_\text{eff}}{\partial\varphi^2}\right|_{\varphi=\varphi_\text{min}}\,=\,2\Lambda^{5}\left(\frac{\rho}{2M_{c}\Lambda^{5}}\right)^{3/2}\,\,\,,
\end{eqnarray}
which clearly scales with $\rho$. In regions of low density, e.g.\,\,the intergalactic vacuum, the chameleon is light, which means it mediates a long range force. In regions of high density, e.g.\,\,in a laboratory, the chameleon becomes massive, which means the force becomes short-ranged, making it challenging to detect with fifth force tests.


\subsection{Symmetrons}

The symmetron model has a conformal coupling and a potential given by \cite{Hinterbichler2010}
\begin{eqnarray}
A^2\left(\varphi\right)&=&\exp\left[\varphi^2/2M_s^2\right]
\,\,\,,
\end{eqnarray}
and
\begin{eqnarray}
V\left(\varphi\right)&=&-\frac{\mu^2}{2}\varphi^2+\frac{\lambda_s}{4}\varphi^4
\,\,\,,
\end{eqnarray}
respectively. As for the chameleon, $M_s$ gives the symmetron-matter coupling and $\lambda_s$ determines the self-interaction strength. Unlike the chameleon, the symmetron effective potential 
\begin{eqnarray}
\label{eq:SymmEffPot}
V_\text{eff}\left(\varphi\right)&=&\frac{1}{2}\left(\frac{\rho}{M_s^2}-\mu_s^2\right)\varphi^2+\frac{\lambda_s}{4}\varphi^4
\end{eqnarray}
has a $\mathbb{Z}_2$ symmetry ($\varphi\rightarrow-\varphi$) which can be spontaneously broken in environments of low matter density, i.e.\,\,when the coefficient of the quadratic term in $\varphi$ is negative. This allows the symmetron to obtain a non-vanishing effective mass in regions where the ambient matter density is below the critical density $\rho^*=\mu^2M_s^2$. The symmetron field has a vanishing vacuum expectation value in high density regions ($\rho>\rho^*$) and thus a vanishing force. Consequently, the parameter $\mu$ determines the scale of the symmetron-matter decoupling.

\begin{figure}[htbp]
\begin{center}
\includegraphics[width=0.45\textwidth]{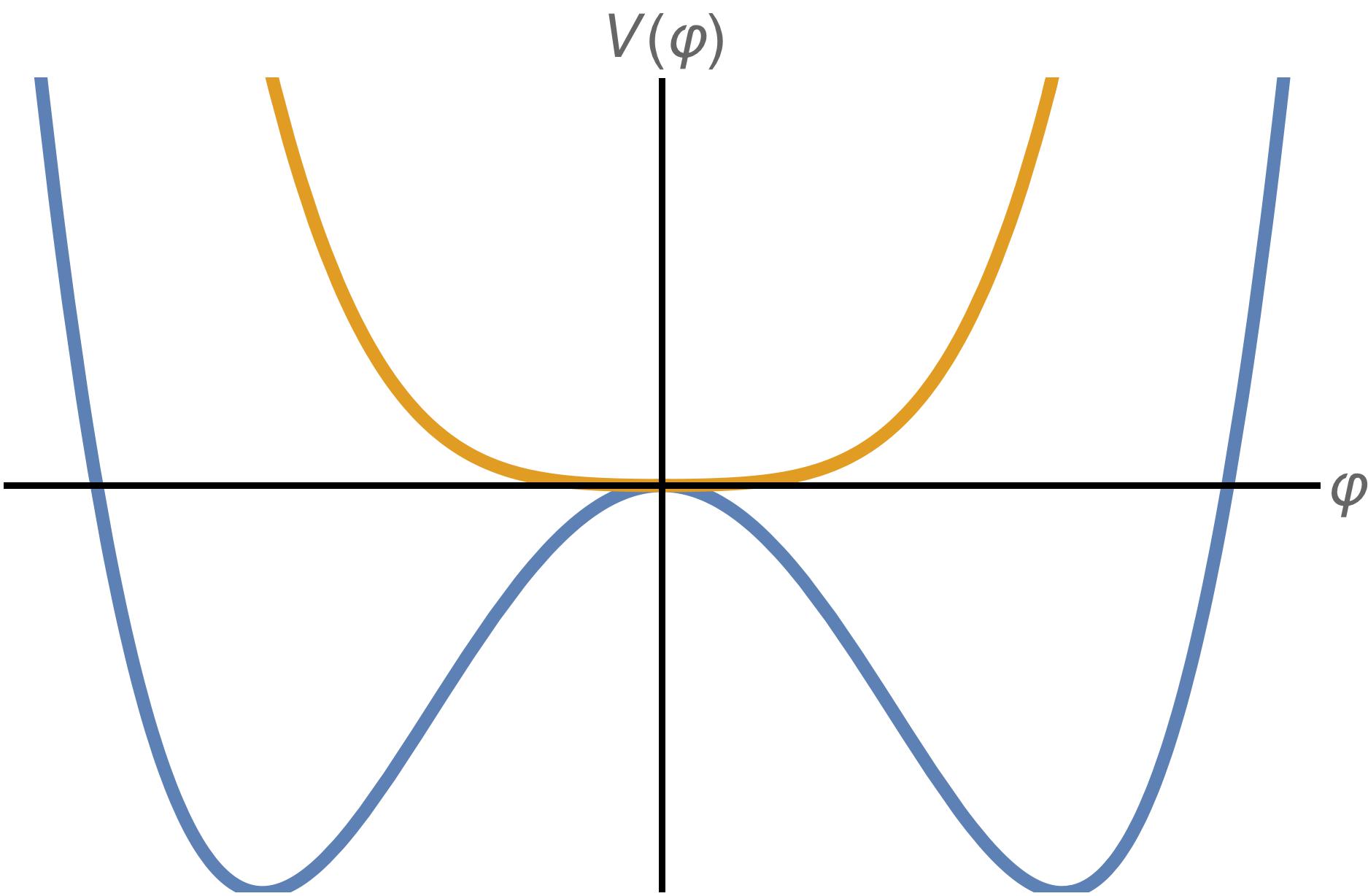}
\caption{Symmetron effective potential, see Eq.\,\,(\ref{eq:SymmEffPot}), for matter densities higher (blue) and lower (orange) than the critical density}
\label{Fig:veff_symmetron}
\end{center}
\end{figure}

Fig.\,\ref{Fig:veff_symmetron} compares the symmetron effective potential for high (orange) and low (blue) density environments. Above the critical density $\rho^*$ (Fig.\,\ref{Fig:veff_symmetron}, orange curve), the minima of $V_\text{eff}$ are degenerate at $\varphi=0$, and so there is no fifth force. Below this density (Fig.\,\ref{Fig:veff_symmetron}, blue curve), the minima become non-degenerate and non-zero at an effective mass of
\begin{eqnarray}
m_s^2&=&2\left(\mu_s^{2}-\frac{\rho}{M_{s}^{2}}\right)\,\,\,.
\end{eqnarray}


\section{BEC interferometer}
\label{sec:BEC}

\begin{figure}[t]
\begin{center}
\includegraphics[width=56mm]{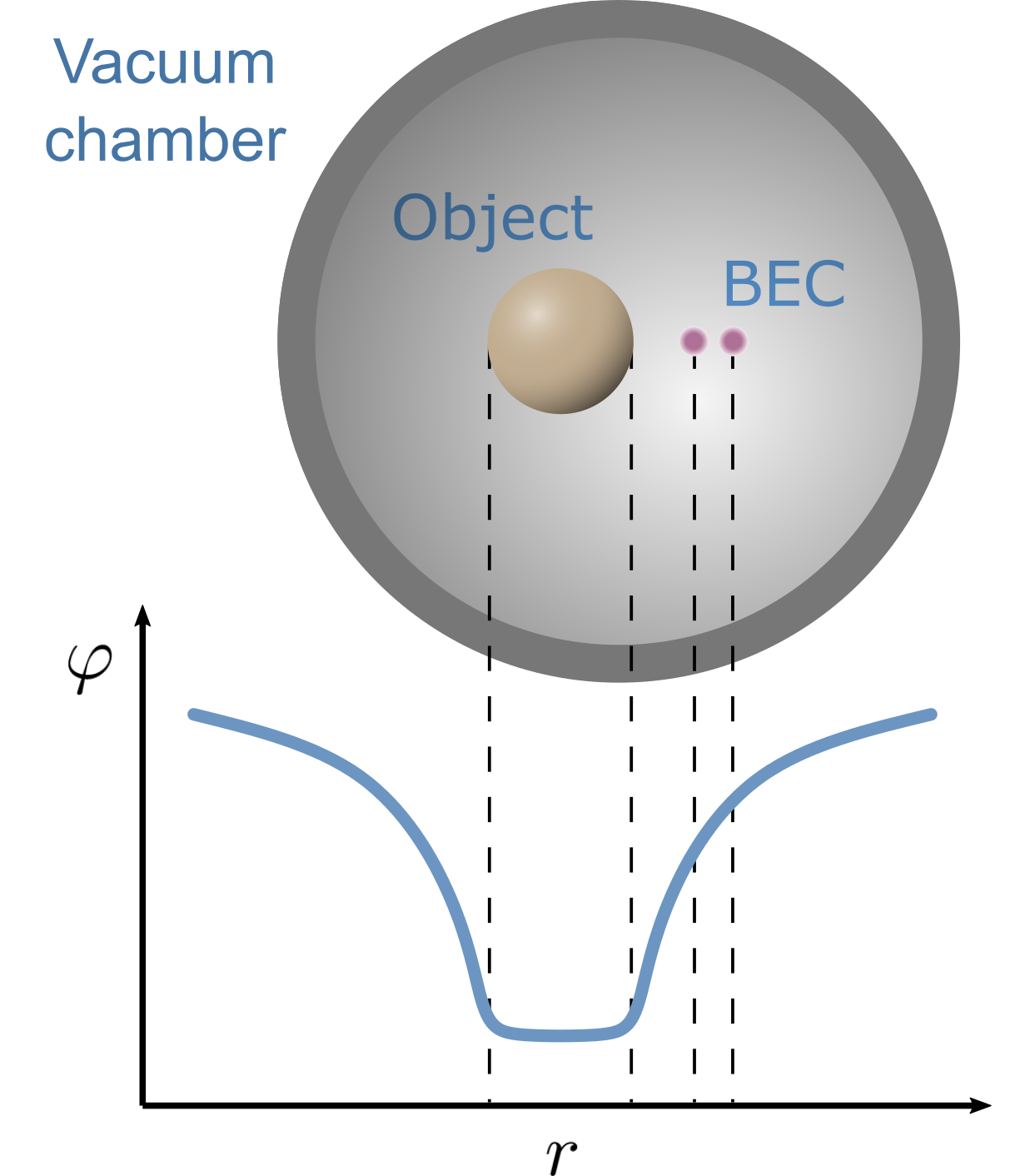}
\caption{A schematic diagram of the vacuum chamber overlaid on the field profile of a chameleon field around a spherical source object; the separation of the BEC components is greatly exaggerated.}
\label{Fig:setup}
\end{center}
\end{figure}

We propose to use a BEC interferometer held near some source mass to constrain the chameleon and symmetron models (Fig.\,\ref{Fig:setup}). Such a scheme would involve trapping bosonic atoms cold enough to form a BEC near a source mass, coherently splitting this cloud into two positions at different radial distances from this source mass, and then interfering the split BEC to extract a phase difference. This phase difference would be used to estimate the potential difference at these two positions, due both to gravity and a potential fifth force. How this might be achieved in practice is discussed in sections \ref{ssec:Impl} and \ref{subsec:numerical_constraints}.

The lowest order gravitational effect of the source mass is a gravitational redshift, which manifests as a position dependent global phase, while the lowest order potential fifth force effect is a modification of this global phase by a position dependent value. This total global phase $\theta$ (derived below) is given by
\begin{eqnarray}\label{eq:GloPhase}
\theta\left(r\right)&=&\frac{mc^{2}T}{2\hbar}\left[\frac{r_{s}}{r}-2\log{A\left(\varphi\left(r\right)\right)}\right]\,\,\,,
\end{eqnarray}
where $r_s$ is the Schwarzschild radius of the source object, $m$ is the mass of each atom in the BEC, $T$ is the time and $A$ is the conformal factor defined as in Sec.\,\ref{sec:Models}. 

\subsection{Global phase}

In what follows, we will motivate the expression of the global phase in Eq.\,(\ref{eq:GloPhase}). For this, we begin by modelling our BEC as an interacting massive scalar Bose field $\hat{\Psi}\left(\boldsymbol{x},t\right)$ in a covariant  formalism to introduce the background metric in a natural way, following the approach of Refs.\,\cite{Fagnocchi2010,Hartley2018a,Hartley2019}. Note that we do not assume that the BEC has relativistic properties such as large excitation energies (i.e.\,\,mass energy), high flow velocities (i.e.\,\,speed of light) or a strong interaction strength etc., and will later explicitly make non-relativistic restrictions.

Following the above references, we describe the evolution of the field operator $\hat{\Psi}$ with the Lagrangian density
\begin{eqnarray}\label{eq:genlag}
\mathcal{L}&=&-\sqrt{-g}\left\{\partial^{\mu}\hat{\Psi}^{\dagger}\partial_{\mu}\hat{\Psi}+\left(\frac{m^{2}c^{2}}{\hbar^{2}}+V\right)\hat{\Psi}^{\dagger}\hat{\Psi}+U\right\}\,\,\,,
\end{eqnarray}
where $V$ is the external potential, $U$ is the interaction potential and $g_{\mu\nu}$ is the metric of the background (in general curved) spacetime with determinant $g$. As is standard in BEC literature \cite{PitStrBEC,PethSmBEC}, we consider only the leading order $2$-particle contact interactions and approximate the interaction strength as
\begin{eqnarray}
U&=&\frac{\lambda}{2}\hat{\Psi}^\dagger\hat{\Psi}^\dagger\hat{\Psi}\hat{\Psi}\,\,\,.
\end{eqnarray}
The interaction strength $\lambda$ can be related to the s-wave scattering length $a_s$ by
\begin{eqnarray}
\lambda=8\pi a_s\,\,\,.
\end{eqnarray}

We can rewrite the field operator $\hat\Psi$ as
\begin{eqnarray}\label{eq:nonrelfield}
\hat\Psi&=&\hat\phi e^{\mathrm{i}mc^2t/\hbar}\,\,\,.
\end{eqnarray}
Later we will make the assumption that time derivatives of $\hat\phi$ are small, i.e.\,\,the excitations described by $\hat\phi$ have non-relativistic energies.

The appropriate background metric near a sphere of radius $R$ and mass $M$ sourcing screening for the assumed screened scalar field has the line element
\begin{eqnarray}\label{eq:lineelement}
ds^2=e^{\zeta^2(r)}\left[-f\left(r\right)dt^2+f^{-1}\left(r\right)dr^2+r^2d\Omega^2\right]
\,\,\,,
\end{eqnarray}
where $f\left(r\right)=1-r_s/r$, $r_s=2GM/c^2$ is the Schwarzschild radius of the object and the conformal factor $A$ has been rewritten as $A^2\left(\varphi\right)=\exp\left[\zeta^2\left(\varphi\right)\right]$ for notational convenience. Eq.\,(\ref{eq:lineelement}) reduces to the Schwarzschild metric when $\zeta^2\rightarrow0$. The gravitational effect of the Earth is ignored; it is assumed that this can be accounted for either with differential measurements with and without the mass, or through a dual interferometer scheme, or simply by splitting the interferometer horizontally.

We now convert this Lagrangian to a Hamiltonian density (for readability and ease of interpretation) and make the following assumptions:
\begin{enumerate}
\item $\left|\zeta^{2}\right|\ll1$\,\,\,,
\item $r_{s}\ll r$\,\,\,,
\item $\left|\partial_{t}\hat{\phi}\right|/c\ll\left|\partial_{i}\hat{\phi}\right|$\,\,\,, and
\item $\hbar^2\left|\partial_i\hat\phi^\dagger\partial_i\hat\phi\right|\ll m^2c^2\hat\phi^\dagger\hat\phi$\,\,\,,
\end{enumerate}
where $i$ runs over spatial indices. To lowest order in $\zeta^2$ and $r_s/r$, the resulting Hamiltonian density is
\begin{eqnarray}\label{eq:hamdens}
\mathcal{H}&=&\frac{\hbar^{2}}{2m}\sum_{i}\partial_{i}\hat{\phi}^{\dagger}\partial_{i}\hat{\phi}+V_{eff}\hat{\phi}^{\dagger}\hat{\phi}+\frac{1}{2}\lambda_{NR}\hat{\phi}^{\dagger}\hat{\phi}^{\dagger}\hat{\phi}\hat{\phi}\,\,\,,
\end{eqnarray}
where
\begin{eqnarray}
V_{eff}&=&V_{NR}+\frac{1}{2}mc^{2}\left[\zeta^{2}-\frac{r_{s}}{r_{avg}}\right]\,\,\,.
\end{eqnarray}

The potentials have been rescaled in the form
\begin{eqnarray}
V_{NR}&=&\frac{V}{2m}\,,\,\lambda_{NR}\,=\,\frac{\lambda}{2m}
\end{eqnarray}
as these are the forms of the external potential and interaction strength that usually appear in the Gross-Pitaevskii equation (GPE) \cite{PitStrBEC,PethSmBEC}. The interaction strength $\lambda_{NR}$ is usually written as $g$ (e.g.\,\,in Ref.\,\cite{PitStrBEC}), but we avoid this notation here to avoid confusion with the background spacetime metric.

We have defined $r_{avg}$ as the mean distance of the BEC from the center of the source mass, and expanded $r_s/r$ as
\begin{eqnarray}
\frac{r_s}{r}&=&\frac{r_s}{r_{avg}+r'}\,=\,\frac{r_s}{r_{avg}}\left(1-\frac{r'}{r_{avg}}+\cdots\right)\,\,\,.
\end{eqnarray}
Note that we can neglect all terms except the first if the BEC trap geometry confines the atom cloud to a region of space where the distance $r_p$ from any point $p$ to the center of the source mass fulfills $\left|r_p-r_{avg}\right|\ll r_{avg}$. If this condition is not met, then different areas of the same cloud of atoms experience a different total phase, adding blur to the final measurement. This condition admits more freedom in the directions perpendicular to the source mass field gradient; for example, the BEC could be trapped with a cigar-shaped trapping potential oriented perpendicularly to the source mass and be brought closer to the source mass than an equivalent homogeneous trap.

The total field $\hat{\phi}$ can be written in terms of momentum eigenmodes as \cite{PitStrBEC}
\begin{eqnarray}
\hat{\phi}\left(\boldsymbol{r},t\right)&=&\left[\Psi_{0}\left(\boldsymbol{r}\right)+\hat{\vartheta}\left(\boldsymbol{r},t\right)\right]e^{-\mathrm{i}\mu t/\hbar}\,\,\,,
\end{eqnarray}
where $\Psi_{0}$ corresponds to the momentum ground state, $\mu$ is the chemical potential and $\hat{\vartheta}$ contains all higher order modes. We make the Bogoliubov approximation and also assume that the excited modes of the field are negligibly occupied. If the potentials $V_{NR}$ and $\lambda_{NR}$ are stationary, then the equation of motion for $\Psi_{0}$ derived from the above Hamiltonian density is
\begin{eqnarray}
\left[-\frac{\hbar^{2}}{2m}\nabla^{2}+V_{eff}-\mu+\lambda_{NR}\left|\Psi_{0}\left(\boldsymbol{r}\right)\right|^{2}\right]\Psi_{0}\left(\boldsymbol{r}\right)&=&0\,\,\,.
\end{eqnarray}
This is the time-independent GPE with the potential replaced by the effective potential $V_{eff}$. Since the screened scalar field contribution to $V_{eff}$ is approximately constant across the width of the BEC, this GPE can be solved by splitting the chemical potential into $\mu=\mu_0+\mu_I$ where $\mu_0$ is the chemical potential when $V_{eff}\rightarrow V_{NR}$. The extra term is then given by
\begin{eqnarray}
\mu_I=\frac{1}{2}mc^{2}\left[\zeta^{2}-\frac{r_{s}}{r_{avg}}\right]\,\,\,.
\end{eqnarray}
Thus, the lowest order effect on the BEC ground state is a shift in the chemical potential, i.e.\,\,a phase shift. Physically, this phase is the gravitational red-shift due to the source mass, and the lowest order contribution of the screened scalar field is a modification of this red-shift. It is also worth noting that this phase shift appears in both the ground state and all excited modes of the BEC in a basis independent way.

\subsection{Phase estimation}

The mean squared error in a specific measurement is bounded from below by the Cram\'{e}r-Rao bound
\begin{eqnarray}\label{eq:QCRB}
\left(\Delta\kappa\right)^2\ge\frac{1}{N F\left(\kappa,\hat{M}\right)}
\,\,\,,
\end{eqnarray}
where $\Delta\kappa$ is the absolute error in estimating the parameter $\kappa$ with some measurement operator $\hat{M}$, and $N$ is the number of measurements performed. The Fisher information $F\left(\kappa,\hat{M}\right)$ can be thought of as the amount of information about $\kappa$ which can be extracted with the measurement $\hat{M}$. The quantum Fisher information (QFI) can be defined as the supremum over all possible measurements
\begin{eqnarray}
H\left(\kappa\right)=\text{sup}_{\hat{M}} F\left(\kappa,\hat{M}\right)
\end{eqnarray}
from which the quantum Cram\'{e}r-Rao bound (QCRB) trivially follows as
\begin{eqnarray}\label{eq:QCRB}
\left(\Delta\kappa\right)^2\ge\frac{1}{N H\left(\kappa\right)}
\,\,\,,
\end{eqnarray}
where $H\left(\kappa\right)$ is the QFI for estimating the parameter $\kappa$. For a full derivation, see \cite{Braunstein1994,Paris2009,Safranek2016}. While the calculation of the Fisher information is generally well defined, it is often difficult to show that a particular measurement is optimal and thus calculate the QFI. Fortunately, this problem has been solved for Gaussian states \cite{Monras2006,Pinel2013,Safranek2015}.

Gaussian states cover the majority of easily experimentally accessible states such as coherent states, thermal states and squeezed states. Calculating the QFI for Gaussian states is simple as Gaussian states have a straightforward description in terms of their first and second moments \cite{Monras2006,Pinel2013,Safranek2015}.

Let $\theta_-$ be the accumulated phase difference between two arms of a BEC interferometer. The QFI for estimating $\theta_-$ with a fully condensed $\mathcal{N}_0$-atom BEC is given by
\begin{eqnarray}
H\left(\theta_-\right)=\mathcal{N}_0
\,\,\,,
\end{eqnarray}
which scales with the standard quantum limit (SQL) \cite{KokLovettQIP}.

\subsection{Implementation}
\label{ssec:Impl}

A BEC coherently split into parts would measure the gradient of the phase in Eq.\,(\ref{eq:GloPhase}) and thus the field gradient in an interference measurement. The other contributions from the environment (e.g.\,\,the gravity of the Earth) could be subtracted with differential measurements or a dual interferometer scheme where measurements are performed near to and far from the source object.

BEC-based interferometers are not a new concept, and have already been proposed and demonstrated (see, e.g.\,\,Refs.\,\cite{Shin2004,Debs2011,Berrada2013,Muntinga2013,McDonald2014}, also Ref.\,\cite{Cronin2009} and references therein). Coherent splitting of a BEC into spatially separated clouds has been implemented both with atom chips \cite{Jo2007,Zhou2016,Berrada2013} (chips printed with an electrode structure allowing for the generation of magnetic and radio-frequency fields very close to an atom cloud) and in various free space arrangements including spatially varying optical dipole traps and optical lattices \cite{Shin2004,McDonald2014,Naik2018,Albiez2005}. Recombination and interference of the separated clouds of a guided atom interferometer is typically achieved by turning the trapping potential off and letting the clouds expand into each other as they fall \cite{Cronin2009}. Note that this is a similar recombination strategy to launched or dropped atom interferometry such as in \cite{Hamilton2015,Jaffe2017}; a guided interferometer confers the additional benefit of potentially greatly extending the interaction time (and thus accumulated phase difference) before recombination. An alternative scheme has been recently realised, where the two condensate parts are brought into contact via Josephson tunnelling through a low potential barrier \cite{Berrada2016}. This acts as a beam splitting operation, and the interference contrast is projected onto a mean atom number difference between the two wells.


\section{Expected bounds}

The expected new bounds on the chameleon and symmetron models from an implementation of our proposed schemes are presented in Figs.\,\ref{Fig:chameleon_n1} - \ref{Fig:symmetron_exclusion}. They are derived from the QCRB for estimating the phase difference in an interferometer. This bound is given by (cf.\,Eq.\,(\ref{eq:QCRB}))
\begin{eqnarray}
\left(\Delta\theta_{-}\right)^{2}\ge\frac{1}{\sqrt{NH\left(\theta_{-}\right)}}\,\,\,.
\end{eqnarray}
Assuming a null measurement, the bounds on the screening models are given by
\begin{eqnarray}
\frac{1}{\sqrt{NH\left(\theta_-\right)}}\ge\frac{mc^2T}{2\hbar}\left(\zeta^2\left(r_1\right)-\zeta^2\left(r_0\right)\right)
\end{eqnarray}
for phases measured at $r_0$ and $r_1$.

We initially give the general form of the relevant bounds for each model in sections \ref{subsec:chameleon_general_constraints} and \ref{subsec:symmetron_general_constraints}, and then present the bounds with some specific numbers from experimental literature in section \ref{subsec:numerical_constraints}.

We note that the analysis we present here is somewhat simplified, for the sake of producing analytic results with straightforward transparent physical justification. Proper analysis of a specific implementation will require numerical analysis including the exact shape of the vacuum chamber, mounting system for the source mass, any additional apparatus required inside the vacuum chamber for trapping the BEC etc. which may enhance or diminish the bounds presented here.

\subsection{Chameleon constraints}\label{subsec:chameleon_general_constraints}

Fig.\,\ref{Fig:chameleon_n1} shows the predicted new constraints for one of the most popular screening models - the chameleon with $n=1$.
There it can be seen that the BEC interferometry scheme would be able to improve existing constraints for this model by up to 3 orders of magnitude and confirm a recent measurement \cite{Yin2022} closing the gap between former interferometry and E\"ot-Wash experiment constraints on the DE scale $\Lambda = 2.4 ~\text{meV}$. This amounts to ruling out the simplest chameleon model as a model of dark energy.

There are four important bounds contributing to the constrained region of the chameleon model parameter space; where the source mass, BEC and residual gas in the vacuum are all screened, where they are all unscreened, where the BEC becomes screened, and where the Compton wavelength of the equilibrium chameleon is larger than the diameter of the vacuum chamber.

In the limit of an infinitely wide vacuum chamber, the $n=1$ chameleon field in the (non-perfect) vacuum has an effective mass
\begin{eqnarray}
m_{\infty}^{2}&=&2\Lambda^{5}\left(\frac{\rho_{\infty}}{2M_{c}\Lambda^{5}}\right)^{3/2}
\,\,\,,
\end{eqnarray}
where $\rho_\infty$ is the matter density of the vacuum. When the chameleon field is screened within the source mass, the constraint resulting from a null measurement is given by
\begin{eqnarray}\label{eq:vacuum_shielded_constraint}
\frac{1}{\sqrt{NH\left(\theta_{-}\right)}}>\frac{mc^{2}T}{2\hbar}\sqrt{\frac{2\Lambda^{5}}{M_{c}}}\left(\frac{1}{\sqrt{\rho_{\infty}}}-\frac{1}{\sqrt{\rho_{obj}}}\right)Re^{m_{\infty}R/\hbar}\left|\frac{e^{-m_{\infty}r_{1}/\hbar}}{r_{1}}-\frac{e^{-m_{\infty}r_{0}/\hbar}}{r_{0}}\right|
\,\,\,,
\end{eqnarray}
where $\rho_{obj}$ is the density of the source object. This expression is only valid if the atoms in the BEC are test particles and do not significantly affect the evolution of the chameleon field profile. However, this is not true for the entire parameter space. As an overly conservative upper bound, we can replace the vacuum matter density $\rho_\infty$ in (\ref{eq:vacuum_shielded_constraint}) with the BEC average matter density $\rho_{BEC}$. The true constraint will lie somewhere between these two, and will require numerical analysis with the specific experimental geometry of any implementation, as in \cite{Copeland2014,Jaffe2017}.

For larger values of $M_c$, the infinite vacuum chamber approximation does not hold as the Compton wavelength of the equilibrium chameleon becomes larger than the size of the vacuum chamber. In this case, the field equilibrium inside the vacuum chamber is instead described by \cite{Elder2016}
\begin{eqnarray}\label{eqn:ChamBackground}
\varphi_\infty\rightarrow\xi\left(n(n+1)\Lambda^{4+n}R^2\right)^{\frac{1}{n+2}}\,\,\,,
\end{eqnarray}
where $\xi=0.55$ is a fudge factor given by the chamber's spherical geometry and vacuum density. The effective mass is set to the radius of the vacuum chamber $m_\infty\rightarrow\hbar/R_\text{vac}$ and the relevant constraint from a null measurement is
\begin{eqnarray}
\frac{1}{\sqrt{NH\left(\theta_{-}\right)}}>\frac{mc^{2}T}{2\hbar}\left(\xi\left[\frac{2\Lambda^{5}R_{\text{vac}}^{2}}{M_{c}^{3}}\right]^{1/3}-\sqrt{\frac{2\Lambda^{5}}{M_{c}\rho_{obj}}}\right)Re^{R/R_{\text{vac}}}\left|\frac{e^{-r_{1}/R_{\text{vac}}}}{r_{1}}-\frac{e^{-r_{0}/R_{\text{vac}}}}{r_{0}}\right|\,\,\,.
\end{eqnarray}

When everything in the vacuum chamber is unscreened, the field profile becomes $\Lambda$-independent and we have
\begin{eqnarray}
\frac{1}{\sqrt{NH\left(\theta_{-}\right)}}>\frac{mc^{2}T}{2\hbar}\left(\frac{\rho_{obj}R^{3}}{3M_{c}^{2}}\right)e^{R/R_{\text{vac}}}\left|\frac{e^{-r_{1}/R_{\text{vac}}}}{r_{1}}-\frac{e^{-r_{0}/R_{\text{vac}}}}{r_{0}}\right|\,\,\,.
\end{eqnarray}

\subsection{Symmetron constraints}\label{subsec:symmetron_general_constraints}

The predicted constraints on the parameter space of the symmetron model are shown in Fig.\,\ref{Fig:symmetron_exclusion}. We expect that our proposed experiment would improve the existing constraints by between 16 and 26 orders of magnitude in $\lambda$ across the entire accessible range of $M_s$.

The value of $\mu_s$ to which these constraints apply is limited by the geometry of the proposed experiment, as the Compton wavelength in low density regions is approximately $1/\mu_s$. For the field to evolve to its vacuum minimum within the chamber, the Compton wavelength must be smaller than the vacuum chamber radius. However, if the Compton wavelength is too small then the field is Yukawa suppressed. We give a numerical estimate for this constraint in section \ref{subsec:numerical_constraints}. 

An object is screened from the symmetron force when its density is above the critical density $\rho^*$. The region in $M_s$ that our proposed experiment would constrain is the region where this critical density is between the densities of the source object and the surrounding vacuum, so
\begin{eqnarray}
\rho_{\infty}<\mu^2M_s^2<\rho_{obj}.
\end{eqnarray}

Finally, the full bound is given by
\begin{eqnarray}
&&\frac{1}{\sqrt{NH\left(\theta_{-}\right)}}>\frac{mc^{2}T}{2\hbar}  \frac{\mu_s^{2}}{\lambda M_{s}^{2}}\left(1-\frac{\rho_{\infty}}{\mu_s^{2}M_{s}^{2}}\right)
\nonumber
\\
&&
\,\,\,\,\,\,\,\,\,\,\,\,\,\,\,\,\,\,\,\,\,\,\,\,\,\,\,\,\,\,\,\,\,\,\,\,\,\,\,\,\,\,\,\,\,\,\,\,\,\,\,\,\,
\times
\left(2\Gamma\left[\frac{e^{-m_{\infty}r_{0}}}{r_{0}}-\frac{e^{-m_{\infty}r_{1}}}{r_{1}}\right]+\Gamma^{2}\left[\frac{e^{-2m_{\infty}r_{1}}}{r_{1}^{2}}-\frac{e^{-2m_{\infty}r_{0}}}{r_{0}^{2}}\right]\right)
\,\,\,,
\end{eqnarray}
where
\begin{eqnarray}
\Gamma&=&R\,e^{m_{\infty}R}\frac{m_{obj}R-\tanh\left(m_{obj}R\right)}{m_{obj}R+m_{\infty}R\tanh\left(m_{obj}R\right)}
\,\,\,,
\end{eqnarray}
and
\begin{eqnarray}
m_{obj/\infty}^{2}&=&2\left(\mu_s^{2}-\frac{\rho_{obj/\infty}}{M_{s}^{2}}\right)\,\,\,.
\end{eqnarray}

\subsection{Numerical constraint estimates}\label{subsec:numerical_constraints}

\begin{figure}[htbp]
\begin{center}
\includegraphics[width=86mm]{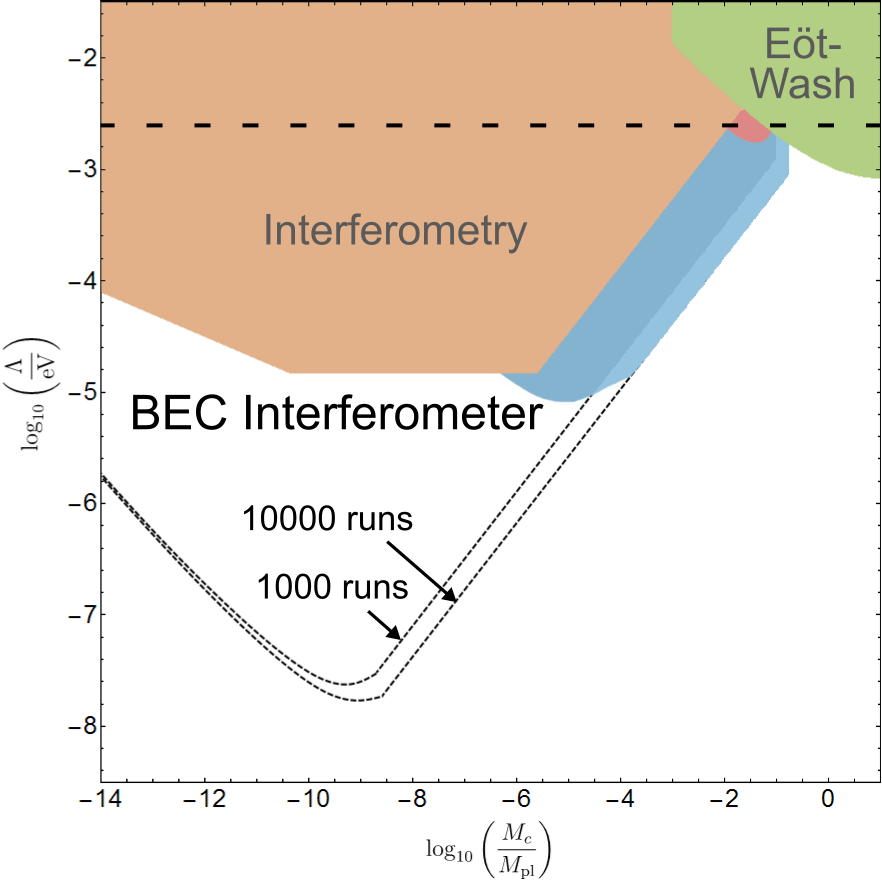}
\caption{Constraints for the parameter space of the chameleon model $n=1$: The brown area corresponds to constraints from atom interferometry, the green area to those from E\"ot-Wash experiments and the red area to recent levitated force sensor results \cite{Sakstein2016,Burrage2017,Yin2022}. The straight dotted line indicates the DE scale $\Lambda = 2.4 ~\text{meV}$. New constraints predicted in this work are coloured in blue, where dark blue corresponds to 1000 runs and light blue corresponds to 10000 runs. The dashed constraints are derived under the assumption that the BEC atoms do not screen the chameleon fifth force, which is implemented by using the vacuum matter density to derive the effective chameleon mass. As the opposite extreme scenario, the solid blue constraints use the average BEC matter density  for deriving the chameleon's mass, i.e., as if the entire vacuum chamber were filled with the BEC. The former case results in too strong constraints, while the latter leads to too weak ones. Consequently, the physically realistic constraints, whose accurate prediction would require the  numerical determination of the chameleon effective mass taking into account each single BEC atom, will lie between the solid blue and the dashed regions.}
\label{Fig:chameleon_n1}
\end{center}
\end{figure}

\begin{figure}[htbp]
\begin{center}
\includegraphics[width=86mm]{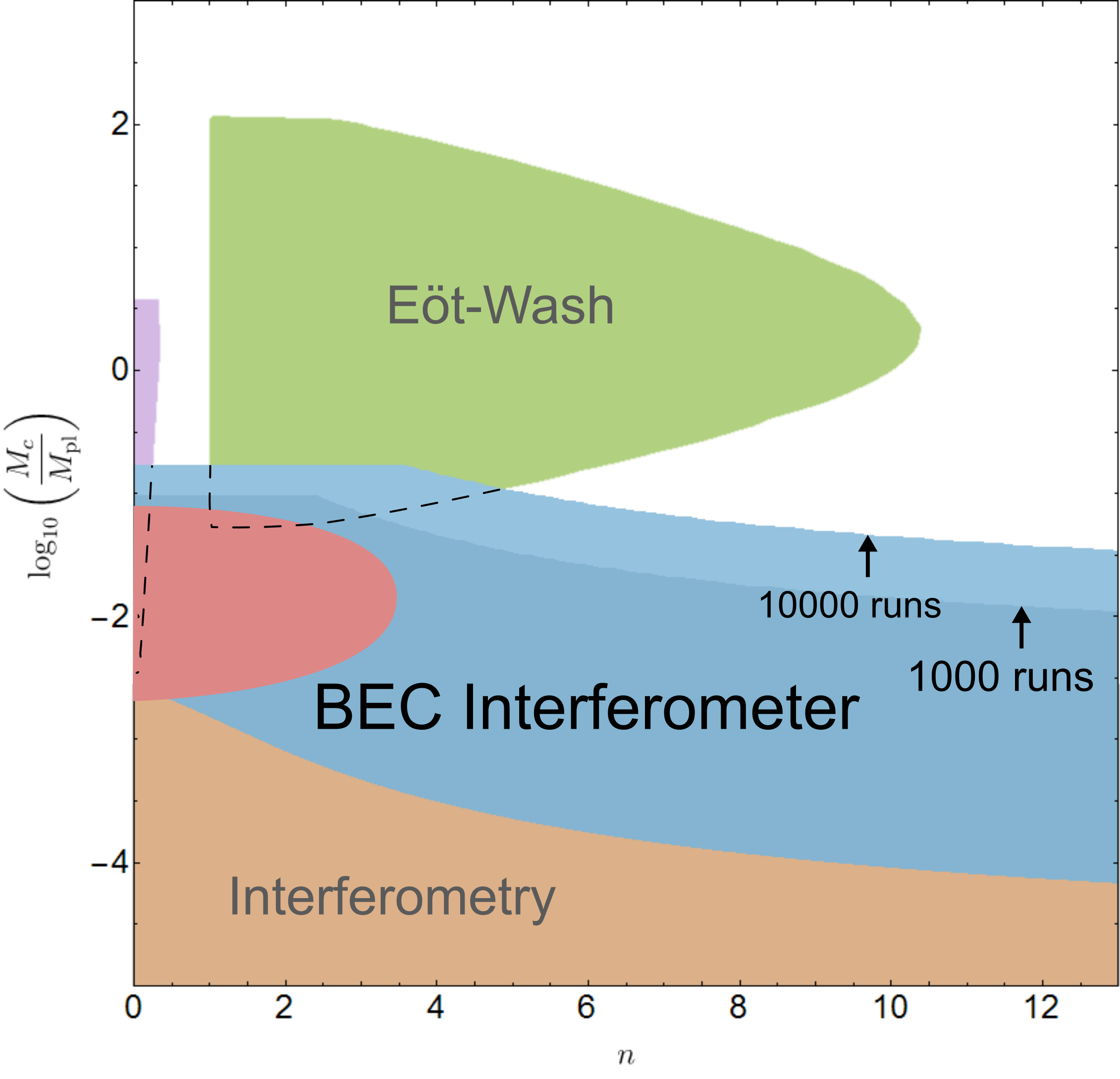}
\caption{Constraints for the value of $M_c$ for positive $n$ chameleon models at $\Lambda = 2.4 ~\text{meV}$: The brown area corresponds to constraints from atom interferometry, the green area to those from E\"ot-Wash experiments, the violet area represents constraints from astrophysics and the red area corresponds to recent levitated force sensor results \cite{Sakstein2016,Burrage2017,Yin2022}.  New constraints predicted in this work are coloured in blue, where dark blue corresponds to 1000 runs and light blue corresponds to 10000 runs.}
\label{Fig:chameleon_varn}
\end{center}
\end{figure}

\begin{figure}[htbp]
\begin{center}
\includegraphics[width=86mm]{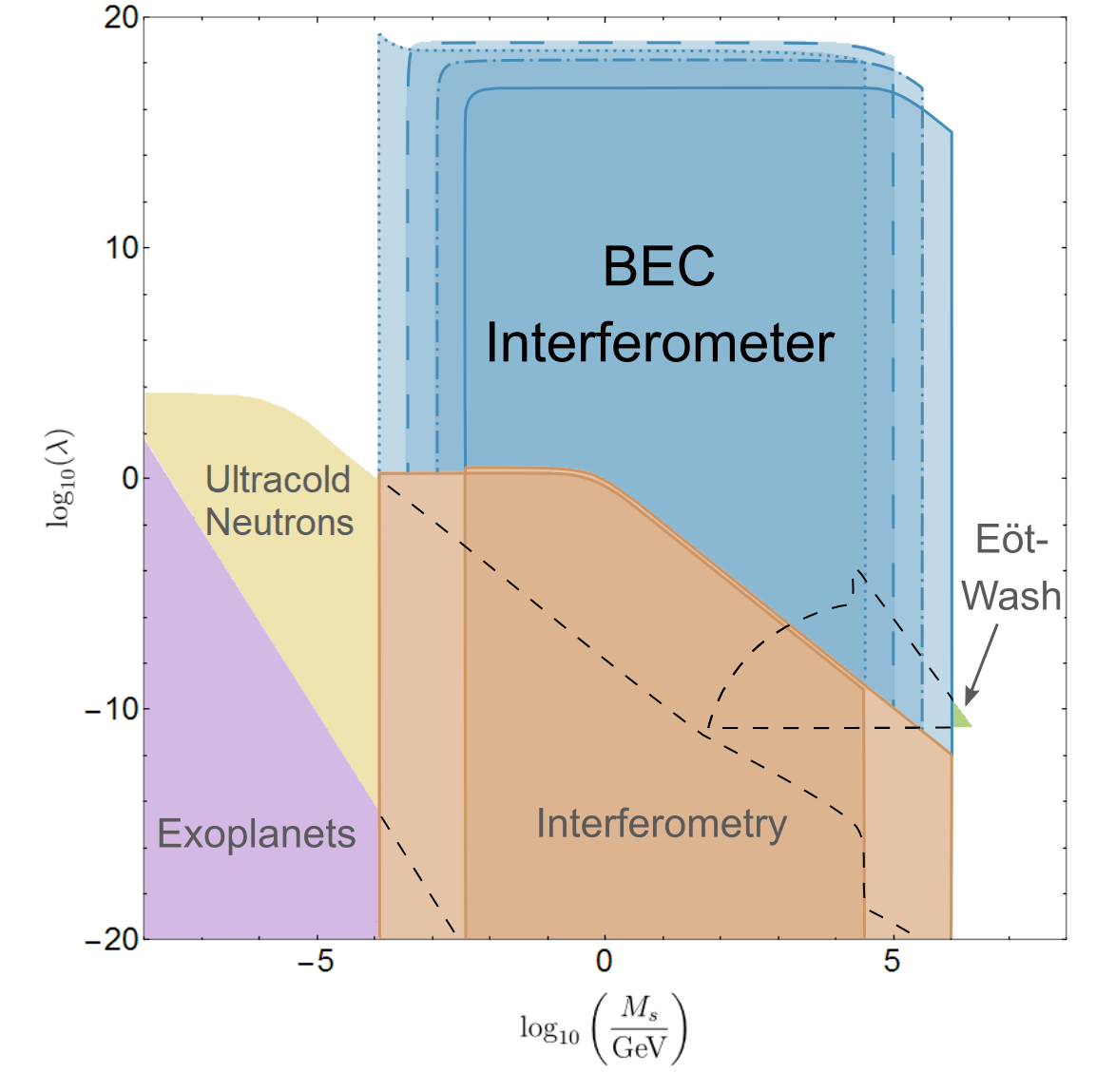}
\caption{Constraints for the parameter space of the symmetron model: The brown area corresponds to constraints from atom interferometry, the green area to those from E\"ot-Wash experiments, and the violet area represents constraints from exoplanet astrophysics \cite{Sakstein2016,Burrage2017}. In additon, the yellow area represents combined constraints from ultracold neutrons for micron- and fermi-screening for $\mu = 10^{-4}$ and $10^{-5}$ eV \cite{Cronenberg2018,Jenke2020}. New constraints predicted in this work for a BEC interferometer are coloured in blue. Differently outlined regions correspond to $\mu=10^{-4}$ (dots), $10^{-4.5}$ (dashes), $10^{-5}$ (mixed dots and dashes) and $10^{-5.5}$ eV (solid) in natural units respectively.}
\label{Fig:symmetron_exclusion}
\end{center}
\end{figure}

We now consider some experimental limitations to the schemes proposed in this article and use these to calculate the expected sensitivity of our schemes to constraining the chameleon and symmetron models. In this section, we consider realistic experimental parameters to demonstrate that a measurement would result in an extension of previously excluded regions of parameter space, but leave a specific implementation to future work.

To numerically estimate the above bounds, we consider the same experimental dimensions as in \cite{Hamilton2015} for ease of comparison. Specifically, we consider a spherical vacuum chamber of radius $L=5$ cm and vacuum pressure $6\times10^{-10}$ Torr. The source object is an aluminium sphere with a radius of $R=9.5$ mm, the effective distance between the object and the BEC is 8.8 mm, and we assume that the two parts of the BEC are split by 100 $\mu$m. With clever trap positioning, the distance between the object and the BEC may eventually be limited by the strength of the van der Waals or Casimir-Polder forces, but these are not relevant at the 10 mm scale.

Typical BEC experiments condense clouds consisting of $10^4-10^6$ atoms, although condensates of up to $10^8$ atoms have been demonstrated with sodium \cite{vdStam2007}, and up to $10^9$ atoms has been demonstrated with hydrogen \cite{Fried1998,Greytak2000}. For estimating the sensitivity of this detector, we assume an initial BEC with $10^6$ atoms, constrained to a quasi-1D trap of length $50~\mu$m.

The maximum integration time of our proposed detector is set by the mutual coherence time of the components of the split BEC. Mutual coherence times up to 500 ms have been demonstrated with atom chips \cite{Jo2007,Zhou2016}, and up to 70s in free space \cite{Panda2022} so we will estimate the integration time of our detector to be 500 ms.

With these numbers, a null measurement would produce the bounds on chameleon models given in Figs.\,\,\ref{Fig:chameleon_n1} for $n=1$ with variable $\Lambda$, and \ref{Fig:chameleon_varn} for $\Lambda=2.4 ~\text{meV}$ for variable $n$.

In Fig.\,\,\ref{Fig:chameleon_n1}, the BEC density bound, i.e.\,\,where we replaced the vacuum matter density $\rho_\infty$ in (\ref{eq:vacuum_shielded_constraint}) with the BEC average matter density $\rho_{BEC}$, defines the blue constraints and the vacuum density bound defines the dashed constraints. We note again that the actual constraint will lie between these two regions, but determining exactly where requires numerical analysis with a specific experimental geometry. With even the most conservative bound, our proposed experiment would cover the gap between former interferometry and E\"ot-Wash experimental constraints on the DE scale $\Lambda = 2.4 ~\text{meV}$.

Fig.\,\ref{Fig:chameleon_varn} shows constraints for the value of $M_c$ over different values of positive $n$ chameleon models and for $\Lambda=\Lambda_\text{DE}=2.4 ~\text{meV}$. The horizontal boundaries in Fig.\,\ref{Fig:chameleon_varn} for early values of $n$ result from the source and the vacuum both being screened.   
For larger values of $n$, the background field profile given in Eq. (\ref{eqn:ChamBackground}) is used. Our scheme would improve existing interferometry constraints by more than 2 orders of magnitude and close the gap to E\"ot-Wash for $n\le5$.

The constraints on the symmetron model are shown in Fig. \ref{Fig:symmetron_exclusion}. As explained in section \ref{subsec:symmetron_general_constraints}, given the vacuum chamber width and distance between the source mass and interferometer, the value of $\mu_s$ that this proposed experiment can constrain is restricted to
\begin{eqnarray}\label{eq:symmetron_mu_bound}
10^{-5.5}\text{ eV}\lesssim\mu_s\lesssim10^{-4}\text{ eV}
\end{eqnarray}
in natural units. The restriction that the critical density must be between the source mass and residual vacuum density causes the sharp sides of the excluded regions in both our predicted excluded regions and the atom interferometry exclusion regions. The curve in the high $M_s$ section of the predicted excluded regions is the area where the critical density and the source object density become comparable, and the peak in the low $M_s$ section of the $\mu_s=10^{-4}$ eV excluded region is caused by a resonance where the Compton wavelength of the symmetron field matches the distance from the object to the BEC.

Wavefunctions of atoms in a BEC are (in the ideal case) spread over the width of the BEC and all overlap. As a first approximation, we consider the BEC to be a region of uniform density as opposed to a collection of discrete objects. The density of the BEC is between that of the vacuum and the source object. When the BEC density is below the critical density, it does not substantially modify the symmetron field profile, but when the BEC density is above the critical density, there should in principle be a dip in the symmetron field profile. However, the Compton wavelength of the symmetron in our constrained region of parameter space given in (\ref{eq:symmetron_mu_bound}) is $\lambda_C\sim10-100$ mm. This is far greater than the width of the BEC in this entire section of the parameter space, so the BEC again does not substantially affect the symmetron field profile. Hence, whether or not the BEC is screened does not play a role in determining the region of parameter space excluded by our proposed experiment.

In the above analysis, we have used experimental numbers, such as the number of atoms and the coherence time, from different physical experiments. We have been conservative with these numbers and expect such an experiment to be achievable. However, to be sure of this, a full analysis of experimental noise and decoherence would need to be taken into account, which is beyond the scope of this work. Such experimental noise would include fluctuations from the trapping potential \cite{PhysRevA.82.063605,PhysRevA.84.043628}, three-body interactions leading to a loss of atoms from the condensate \cite{PhysRevLett.89.140402}, interactions with the thermal cloud \cite{schelle2009environment}, and interactions of condensate atoms with foreign atoms due to an imprecise vacuum \cite{PhysRevA.60.R29,PhysRevA.81.013620}. These sources of noise would also contribute  to the achievable sensitivity of the detector. For example, as atoms are lost from the condensate over time, the  sensitivity of the detector will decrease. Furthermore, interactions with the environment, such as interactions with the thermal cloud, will lead to a loss of coherence in the condensate, also contributing to a reduction in the sensitivity. We leave such detailed analysis to future work.


\section{Conclusions}

Conformally coupled scalar fields like chameleons or symmetrons are popular means for explaining the nature of dark energy. In recent years, various experiments have been performed in order to constrain these models. Some of the most successful experiments are based on cold atoms, e.g.\,\,atom interferometry.  

In this article we proposed a BEC interferometer as a novel way of searching for screened scalar fields, which we expect to lead to massively improved constraints for chameleon and symmetron models. To bring this proposal into reality, future work will focus on optimising the experimental implementation. Any subsequent implementation of our proposal will either discover $n=1$ chameleon fields at the cosmological energy density or confirm completely ruling them out, along with greatly improving the bounds on other screened scalar models.

While we have only considered chameleon and symmetron screening models, it should be stressed that constraints for any other type of conformally coupled scalar field could be obtained in a similar manner, e.g.\,\,for dilatons \cite{Damour1994,Brax2011}.


\begin{acknowledgments}
The authors thank  A. L. B\'{a}ez-Camargo, C. Burrage, B. Elder, P. Millington and M. Pitschmann for helpful comments, C. Burrage and J. Sakstein for providing their constraint plot files, and H. Abele and T. Jenke for providing their symmetron constraints. The authors acknowledge financial support from the Austrian Science Fund (FWF) through project codes W 1210-N25 and P 34240-N, the University of Nottingham, and the John Templeton Foundation through grant No. 58745. RH also acknowledges the John Templeton Foundation grant No. 62312. IF acknowledges financial support from J. Moussouris and E. Jhong. The opinions expressed in this publication are those of the authors and do not necessarily reflect the views of the John Templeton Foundation.
\end{acknowledgments}

\bibliography{Refs}

\providecommand{\href}[2]{#2}\begingroup\raggedright\begin{thebibliography}{10}

\bibitem{Brans1961}
C.~Brans and R.~H. Dicke, \emph{{Mach's Principle and a Relativistic Theory of
  Gravitation}}, \href{https://doi.org/10.1103/PhysRev.124.925}{\emph{Phys.
  Rev.} {\bfseries 124} (1961) 925}.

\bibitem{Fujii2003}
Y.~Fujii and K.-i. Maeda, \emph{The Scalar-Tensor Theory of Gravitation},
  Cambridge Monographs on Mathematical Physics. Cambridge University Press,
  2003,
  \href{https://doi.org/10.1017/CBO9780511535093}{10.1017/CBO9780511535093}.

\bibitem{Wehus2002}
I.~K. Wehus and F.~Ravndal, \emph{{DYNAMICS OF THE SCALAR FIELD IN
  FIVE-DIMENSIONAL KALUZA–KLEIN THEORY}},
  \href{https://doi.org/10.1142/S0217751X04020609}{\emph{International Journal
  of Modern Physics A} {\bfseries 19} (2004) 4671}.

\bibitem{Perlmutter1998}
S.~Perlmutter, G.~Aldering, G.~Goldhaber, R.~A. Knop, P.~Nugent, P.~G. Castro
  et~al., \emph{{Measurements of $\Omega$ and $\Lambda$ from 42 High-Redshift
  Supernovae}}, \href{https://doi.org/10.1086/307221}{\emph{The Astrophysical
  Journal} {\bfseries 517} (1999) 565}.

\bibitem{Riess1998}
A.~G. Riess, A.~V. Filippenko, P.~Challis, A.~Clocchiatti, A.~Diercks, P.~M.
  Garnavich et~al., \emph{{Observational Evidence from Supernovae for an
  Accelerating Universe and a Cosmological Constant}},
  \href{https://doi.org/10.1086/300499}{\emph{The Astronomical Journal}
  {\bfseries 116} (1998) 1009}.

\bibitem{Clifton2011}
T.~Clifton, P.~G. Ferreira, A.~Padilla and C.~Skordis, \emph{Modified gravity
  and cosmology},
  \href{https://doi.org/https://doi.org/10.1016/j.physrep.2012.01.001}{\emph{Physics
  Reports} {\bfseries 513} (2012) 1}.

\bibitem{Joyce2014}
A.~Joyce, B.~Jain, J.~Khoury and M.~Trodden, \emph{{Beyond the Cosmological
  Standard Model}},
  \href{https://doi.org/10.1016/j.physrep.2014.12.002}{\emph{Phys. Rept.}
  {\bfseries 568} (2015) 1} [\href{https://arxiv.org/abs/1407.0059}{{\ttfamily
  1407.0059}}].

\bibitem{Dickey1994}
J.~O. Dickey, P.~L. Bender, J.~E. Faller, X.~X. Newhall, R.~L. Ricklefs, J.~G.
  Ries et~al., \emph{{Lunar Laser Ranging: A Continuing Legacy of the Apollo
  Program}}, \href{https://doi.org/10.1126/science.265.5171.482}{\emph{Science}
  {\bfseries 265} (1994) 482}.

\bibitem{Adelberger2003}
E.~Adelberger, B.~Heckel and A.~Nelson, \emph{{Tests of the Gravitational
  Inverse-Square Law}},
  \href{https://doi.org/10.1146/annurev.nucl.53.041002.110503}{\emph{Annual
  Review of Nuclear and Particle Science} {\bfseries 53} (2003) 77}.

\bibitem{Kapner2007}
D.~J. Kapner, T.~S. Cook, E.~G. Adelberger, J.~H. Gundlach, B.~R. Heckel, C.~D.
  Hoyle et~al., \emph{{Tests of the Gravitational Inverse-Square Law below the
  Dark-Energy Length Scale}},
  \href{https://doi.org/10.1103/PhysRevLett.98.021101}{\emph{Phys. Rev. Lett.}
  {\bfseries 98} (2007) 021101}.

\bibitem{Ishak2018}
M.~Ishak, \emph{{Testing General Relativity in Cosmology}},
  \href{https://doi.org/10.1007/s41114-018-0017-4}{\emph{Living Rev. Rel.}
  {\bfseries 22} (2019) 1} [\href{https://arxiv.org/abs/1806.10122}{{\ttfamily
  1806.10122}}].

\bibitem{Burrage2017}
C.~Burrage and J.~Sakstein, \emph{{Tests of Chameleon Gravity}},
  \href{https://doi.org/10.1007/s41114-018-0011-x}{\emph{Living Rev. Rel.}
  {\bfseries 21} (2018) 1} [\href{https://arxiv.org/abs/1709.09071}{{\ttfamily
  1709.09071}}].

\bibitem{Ludlow2015}
A.~D. Ludlow, M.~M. Boyd, J.~Ye, E.~Peik and P.~O. Schmidt, \emph{Optical
  atomic clocks}, \href{https://doi.org/10.1103/RevModPhys.87.637}{\emph{Rev.
  Mod. Phys.} {\bfseries 87} (2015) 637}.

\bibitem{Schlippert2015}
D.~Schlippert et~al., \emph{{Ground Tests of Einstein's Equivalence Principle:
  From Lab-based to 10-m Atomic Fountains}},  in \emph{{50th Rencontres de
  Moriond on Gravitation: 100 years after GR}}, pp.~153--161, 7, 2015,
  \href{https://arxiv.org/abs/1507.05820}{{\ttfamily 1507.05820}}.

\bibitem{Overstreet2018}
C.~Overstreet, P.~Asenbaum, T.~Kovachy, R.~Notermans, J.~M. Hogan and M.~A.
  Kasevich, \emph{{Effective Inertial Frame in an Atom Interferometric Test of
  the Equivalence Principle}},
  \href{https://doi.org/10.1103/PhysRevLett.120.183604}{\emph{Phys. Rev. Lett.}
  {\bfseries 120} (2018) 183604}.

\bibitem{Becker2018}
D.~Becker, M.~D. Lachmann, S.~T. Seidel, H.~Ahlers, A.~N. Dinkelaker, J.~Grosse
  et~al., \emph{{Space-borne Bose{\textendash}Einstein condensation for
  precision interferometry}},
  \href{https://doi.org/10.1038/s41586-018-0605-1}{\emph{Nature} {\bfseries
  562} (2018) 391}.

\bibitem{Copeland2014}
C.~Burrage, E.~J. Copeland and E.~A. Hinds, \emph{{Probing Dark Energy with
  Atom Interferometry}},
  \href{https://doi.org/10.1088/1475-7516/2015/03/042}{\emph{JCAP} {\bfseries
  03} (2015) 042} [\href{https://arxiv.org/abs/1408.1409}{{\ttfamily
  1408.1409}}].

\bibitem{Jaffe2017}
M.~Jaffe, P.~Haslinger, V.~Xu, P.~Hamilton, A.~Upadhye, B.~Elder et~al.,
  \emph{{Testing sub-gravitational forces on atoms from a miniature, in-vacuum
  source mass}}, \href{https://doi.org/10.1038/nphys4189}{\emph{Nature Phys.}
  {\bfseries 13} (2017) 938}
  [\href{https://arxiv.org/abs/1612.05171}{{\ttfamily 1612.05171}}].

\bibitem{Hartley2019}
D.~Hartley, C.~K\"ading, R.~Howl and I.~Fuentes, \emph{{Quantum simulation of
  dark energy candidates}},
  \href{https://doi.org/10.1103/PhysRevD.99.105002}{\emph{Phys. Rev. D}
  {\bfseries 99} (2019) 105002}
  [\href{https://arxiv.org/abs/1811.06927}{{\ttfamily 1811.06927}}].

\bibitem{Kading1}
C.~Burrage, C.~K\"ading, P.~Millington and J.~Min\'a\v{r}, \emph{{Open quantum
  dynamics induced by light scalar fields}},
  \href{https://doi.org/10.1103/PhysRevD.100.076003}{\emph{Phys. Rev. D}
  {\bfseries 100} (2019) 076003}
  [\href{https://arxiv.org/abs/1812.08760}{{\ttfamily 1812.08760}}].

\bibitem{Kading2}
C.~Burrage, C.~K\"ading, P.~Millington and J.~Min\'a\v{r}, \emph{{Influence
  functionals, decoherence and conformally coupled scalars}},
  \href{https://doi.org/10.1088/1742-6596/1275/1/012041}{\emph{J. Phys. Conf.
  Ser.} {\bfseries 1275} (2019) 012041}
  [\href{https://arxiv.org/abs/1902.09607}{{\ttfamily 1902.09607}}].

\bibitem{Khoury2003}
J.~Khoury and A.~Weltman, \emph{{Chameleon cosmology}},
  \href{https://doi.org/10.1103/PhysRevD.69.044026}{\emph{Phys. Rev. D}
  {\bfseries 69} (2004) 044026}
  [\href{https://arxiv.org/abs/astro-ph/0309411}{{\ttfamily
  astro-ph/0309411}}].

\bibitem{Khoury20032}
J.~Khoury and A.~Weltman, \emph{{Chameleon fields: Awaiting surprises for tests
  of gravity in space}},
  \href{https://doi.org/10.1103/PhysRevLett.93.171104}{\emph{Phys. Rev. Lett.}
  {\bfseries 93} (2004) 171104}
  [\href{https://arxiv.org/abs/astro-ph/0309300}{{\ttfamily
  astro-ph/0309300}}].

\bibitem{Dehnen1992}
H.~Dehnen, H.~Frommert and F.~Ghaboussi, \emph{{Higgs field and a new scalar -
  tensor theory of gravity}},
  \href{https://doi.org/10.1007/BF00674344}{\emph{Int. J. Theor. Phys.}
  {\bfseries 31} (1992) 109}.

\bibitem{Gessner1992}
E.~Gessner, \emph{{A new scalar tensor theory for gravity and the flat rotation
  curves of spiral galaxies}},
  \href{https://doi.org/10.1007/BF00645239}{\emph{Astrophys. Space Sci.}
  {\bfseries 196} (1992) 29}.

\bibitem{Damour1994}
T.~Damour and A.~M. Polyakov, \emph{{The String dilaton and a least coupling
  principle}}, \href{https://doi.org/10.1016/0550-3213(94)90143-0}{\emph{Nucl.
  Phys. B} {\bfseries 423} (1994) 532}
  [\href{https://arxiv.org/abs/hep-th/9401069}{{\ttfamily hep-th/9401069}}].

\bibitem{Pietroni2005}
M.~Pietroni, \emph{Dark energy condensation},
  \href{https://doi.org/10.1103/PhysRevD.72.043535}{\emph{Phys. Rev. D}
  {\bfseries 72} (2005) 043535}.

\bibitem{Olive2008}
K.~A. Olive and M.~Pospelov, \emph{Environmental dependence of masses and
  coupling constants},
  \href{https://doi.org/10.1103/PhysRevD.77.043524}{\emph{Phys. Rev. D}
  {\bfseries 77} (2008) 043524}.

\bibitem{Brax2010}
P.~Brax, C.~van~de Bruck, A.-C. Davis and D.~Shaw, \emph{Dilaton and modified
  gravity}, \href{https://doi.org/10.1103/PhysRevD.82.063519}{\emph{Phys. Rev.
  D} {\bfseries 82} (2010) 063519}.

\bibitem{Hinterbichler2010}
K.~Hinterbichler and J.~Khoury, \emph{{Symmetron Fields: Screening Long-Range
  Forces Through Local Symmetry Restoration}},
  \href{https://doi.org/10.1103/PhysRevLett.104.231301}{\emph{Phys. Rev. Lett.}
  {\bfseries 104} (2010) 231301}
  [\href{https://arxiv.org/abs/1001.4525}{{\ttfamily 1001.4525}}].

\bibitem{Hinterbichler2011}
K.~Hinterbichler, J.~Khoury, A.~Levy and A.~Matas, \emph{{Symmetron
  Cosmology}}, \href{https://doi.org/10.1103/PhysRevD.84.103521}{\emph{Phys.
  Rev. D} {\bfseries 84} (2011) 103521}
  [\href{https://arxiv.org/abs/1107.2112}{{\ttfamily 1107.2112}}].

\bibitem{Burrage2016}
C.~Burrage, A.~Kuribayashi-Coleman, J.~Stevenson and B.~Thrussell,
  \emph{{Constraining symmetron fields with atom interferometry}},
  \href{https://doi.org/10.1088/1475-7516/2016/12/041}{\emph{JCAP} {\bfseries
  12} (2016) 041} [\href{https://arxiv.org/abs/1609.09275}{{\ttfamily
  1609.09275}}].

\bibitem{Nowakowski:2000zd}
M.~Nowakowski, \emph{{Long range forces from quantum field theory at zero and
  finite temperature}}, \href{https://doi.org/10.22323/1.005.0025}{\emph{PoS}
  {\bfseries silafae-III} (2000) 025}
  [\href{https://arxiv.org/abs/hep-ph/0009157}{{\ttfamily hep-ph/0009157}}].

\bibitem{Fagnocchi2010}
S.~Fagnocchi, S.~Finazzi, S.~Liberati, M.~Kormos and A.~Trombettoni,
  \emph{{Relativistic Bose–Einstein condensates: a new system for analogue
  models of gravity}},
  \href{https://doi.org/10.1088/1367-2630/12/9/095012}{\emph{New Journal of
  Physics} {\bfseries 12} (2010) 095012}.

\bibitem{Hartley2018a}
D.~Hartley, T.~Bravo, D.~R\"atzel, R.~Howl and I.~Fuentes, \emph{{Analogue
  simulation of gravitational waves in a $3+1$-dimensional Bose-Einstein
  condensate}}, \href{https://doi.org/10.1103/PhysRevD.98.025011}{\emph{Phys.
  Rev. D} {\bfseries 98} (2018) 025011}.

\bibitem{PitStrBEC}
L.~Pitaevskii and S.~Stringari, \emph{Bose-Einstein Condensation}. Oxford
  University Press, 2003.

\bibitem{PethSmBEC}
C.~J. Pethick and H.~Smith, \emph{Bose–Einstein Condensation in Dilute
  Gases}. Cambridge University Press, 2001,
  \href{https://doi.org/10.1017/CBO9780511755583}{10.1017/CBO9780511755583}.

\bibitem{Braunstein1994}
S.~L. Braunstein and C.~M. Caves, \emph{Statistical distance and the geometry
  of quantum states},
  \href{https://doi.org/10.1103/PhysRevLett.72.3439}{\emph{Phys. Rev. Lett.}
  {\bfseries 72} (1994) 3439}.

\bibitem{Paris2009}
M.~G.~A. Paris, \emph{Quantum estimation for quantum technology},
  \href{https://doi.org/10.1142/S0219749909004839}{\emph{International Journal
  of Quantum Information} {\bfseries 07} (2009) 125}.

\bibitem{Safranek2016}
D.~\ifmmode~\check{S}\else \v{S}\fi{}afr\'anek and I.~Fuentes, \emph{{Optimal
  probe states for the estimation of Gaussian unitary channels}},
  \href{https://doi.org/10.1103/PhysRevA.94.062313}{\emph{Phys. Rev. A}
  {\bfseries 94} (2016) 062313}.

\bibitem{Monras2006}
A.~Monras, \emph{{Optimal phase measurements with pure Gaussian states}},
  \href{https://doi.org/10.1103/PhysRevA.73.033821}{\emph{Phys. Rev. A}
  {\bfseries 73} (2006) 033821}.

\bibitem{Pinel2013}
O.~Pinel, P.~Jian, N.~Treps, C.~Fabre and D.~Braun, \emph{{Quantum parameter
  estimation using general single-mode Gaussian states}},
  \href{https://doi.org/10.1103/PhysRevA.88.040102}{\emph{Phys. Rev. A}
  {\bfseries 88} (2013) 040102}.

\bibitem{Safranek2015}
D.~\ifmmode~\check{S}\else \v{S}\fi{}afr\'anek, A.~R. Lee and I.~Fuentes,
  \emph{Quantum parameter estimation using multi-mode gaussian states},
  \href{https://doi.org/10.1088/1367-2630/17/7/073016}{\emph{New Journal of
  Physics} {\bfseries 17} (2015) 073016}.

\bibitem{KokLovettQIP}
P.~Kok and B.~W. Lovett, \emph{Introduction to Optical Quantum Information
  Processing}. Cambridge University Press, 2010,
  \href{https://doi.org/10.1017/CBO9781139193658}{10.1017/CBO9781139193658}.

\bibitem{Shin2004}
Y.~Shin, M.~Saba, T.~A. Pasquini, W.~Ketterle, D.~E. Pritchard and A.~E.
  Leanhardt, \emph{{Atom Interferometry with Bose-Einstein Condensates in a
  Double-Well Potential}},
  \href{https://doi.org/10.1103/PhysRevLett.92.050405}{\emph{Phys. Rev. Lett.}
  {\bfseries 92} (2004) 050405}.

\bibitem{Debs2011}
J.~E. Debs, P.~A. Altin, T.~H. Barter, D.~D\"oring, G.~R. Dennis, G.~McDonald
  et~al., \emph{{Cold-atom gravimetry with a Bose-Einstein condensate}},
  \href{https://doi.org/10.1103/PhysRevA.84.033610}{\emph{Phys. Rev. A}
  {\bfseries 84} (2011) 033610}.

\bibitem{Berrada2013}
T.~Berrada, S.~van Frank, R.~Bücker, T.~Schumm, J.-F. Schaff and
  J.~Schmiedmayer, \emph{{Integrated Mach{\textendash}Zehnder interferometer
  for Bose{\textendash}Einstein condensates}},
  \href{https://doi.org/10.1038/ncomms3077}{\emph{Nature Communications}
  {\bfseries 4} (2013) }.

\bibitem{Muntinga2013}
H.~M\"untinga, H.~Ahlers, M.~Krutzik, A.~Wenzlawski, S.~Arnold, D.~Becker
  et~al., \emph{{Interferometry with Bose-Einstein Condensates in
  Microgravity}},
  \href{https://doi.org/10.1103/PhysRevLett.110.093602}{\emph{Phys. Rev. Lett.}
  {\bfseries 110} (2013) 093602}.

\bibitem{McDonald2014}
G.~D. McDonald, C.~C.~N. Kuhn, K.~S. Hardman, S.~Bennetts, P.~J. Everitt, P.~A.
  Altin et~al., \emph{{Bright Solitonic Matter-Wave Interferometer}},
  \href{https://doi.org/10.1103/PhysRevLett.113.013002}{\emph{Phys. Rev. Lett.}
  {\bfseries 113} (2014) 013002}.

\bibitem{Cronin2009}
A.~D. Cronin, J.~Schmiedmayer and D.~E. Pritchard, \emph{{Optics and
  interferometry with atoms and molecules}},
  \href{https://doi.org/10.1103/RevModPhys.81.1051}{\emph{Rev. Mod. Phys.}
  {\bfseries 81} (2009) 1051}.

\bibitem{Jo2007}
G.-B. Jo, Y.~Shin, S.~Will, T.~A. Pasquini, M.~Saba, W.~Ketterle et~al.,
  \emph{{Long Phase Coherence Time and Number Squeezing of Two Bose-Einstein
  Condensates on an Atom Chip}},
  \href{https://doi.org/10.1103/PhysRevLett.98.030407}{\emph{Phys. Rev. Lett.}
  {\bfseries 98} (2007) 030407}.

\bibitem{Zhou2016}
S.~Zhou, D.~Groswasser, M.~Keil, Y.~Japha and R.~Folman, \emph{Robust spatial
  coherence $5\phantom{\rule{0.16em}{0ex}}\ensuremath{\mu}\mathrm{m}$ from a
  room-temperature atom chip},
  \href{https://doi.org/10.1103/PhysRevA.93.063615}{\emph{Phys. Rev. A}
  {\bfseries 93} (2016) 063615}.

\bibitem{Naik2018}
D.~S. Naik, G.~Kuyumjyan, D.~Pandey, P.~Bouyer and A.~Bertoldi,
  \emph{{Bose–Einstein condensate array in a malleable optical trap formed in
  a traveling wave cavity}},
  \href{https://doi.org/10.1088/2058-9565/aad48e}{\emph{Quantum Science and
  Technology} {\bfseries 3} (2018) 045009}.

\bibitem{Albiez2005}
M.~Albiez, R.~Gati, J.~F\"olling, S.~Hunsmann, M.~Cristiani and M.~K.
  Oberthaler, \emph{{Direct Observation of Tunneling and Nonlinear
  Self-Trapping in a Single Bosonic Josephson Junction}},
  \href{https://doi.org/10.1103/PhysRevLett.95.010402}{\emph{Phys. Rev. Lett.}
  {\bfseries 95} (2005) 010402}.

\bibitem{Hamilton2015}
P.~Hamilton, M.~Jaffe, P.~Haslinger, Q.~Simmons, H.~Müller and J.~Khoury,
  \emph{Atom-interferometry constraints on dark energy},
  \href{https://doi.org/10.1126/science.aaa8883}{\emph{Science} {\bfseries 349}
  (2015) 849}.

\bibitem{Berrada2016}
T.~Berrada, S.~van Frank, R.~B\"ucker, T.~Schumm, J.-F. Schaff, J.~Schmiedmayer
  et~al., \emph{{Matter-wave recombiners for trapped Bose-Einstein
  condensates}}, \href{https://doi.org/10.1103/PhysRevA.93.063620}{\emph{Phys.
  Rev. A} {\bfseries 93} (2016) 063620}.

\bibitem{Yin2022}
P.~Yin, R.~Li, C.~Yin, X.~Xu, X.~Bian, H.~Xie et~al., \emph{{Experiments with
  levitated force sensor challenge theories of dark energy}},
  \href{https://doi.org/10.1038/s41567-022-01706-9}{\emph{Nature Physics}
  {\bfseries 18} (2022) 1181}.

\bibitem{Elder2016}
B.~Elder, J.~Khoury, P.~Haslinger, M.~Jaffe, H.~M\"uller and P.~Hamilton,
  \emph{{Chameleon Dark Energy and Atom Interferometry}},
  \href{https://doi.org/10.1103/PhysRevD.94.044051}{\emph{Phys. Rev. D}
  {\bfseries 94} (2016) 044051}
  [\href{https://arxiv.org/abs/1603.06587}{{\ttfamily 1603.06587}}].

\bibitem{Sakstein2016}
C.~Burrage and J.~Sakstein, \emph{A compendium of chameleon constraints},
  \href{https://doi.org/10.1088/1475-7516/2016/11/045}{\emph{Journal of
  Cosmology and Astroparticle Physics} {\bfseries 2016} (2016) 045}.

\bibitem{Cronenberg2018}
G.~Cronenberg, P.~Brax, H.~Filter, P.~Geltenbort, T.~Jenke, G.~Pignol et~al.,
  \emph{{Acoustic Rabi oscillations between gravitational quantum states and
  impact on symmetron dark energy}},
  \href{https://doi.org/10.1038/s41567-018-0205-x}{\emph{Nature Phys.}
  {\bfseries 14} (2018) 1022}
  [\href{https://arxiv.org/abs/1902.08775}{{\ttfamily 1902.08775}}].

\bibitem{Jenke2020}
T.~Jenke, J.~Bosina, J.~Micko, M.~Pitschmann, R.~Sedmik and H.~Abele,
  \emph{{Gravity resonance spectroscopy and dark energy symmetron fields:
  qBOUNCE experiments performed with Rabi and Ramsey spectroscopy}},
  \href{https://doi.org/10.1140/epjs/s11734-021-00088-y}{\emph{Eur. Phys. J.
  ST} {\bfseries 230} (2021) 1131}
  [\href{https://arxiv.org/abs/2012.07472}{{\ttfamily 2012.07472}}].

\bibitem{vdStam2007}
K.~M.~R. van~der Stam, E.~D. van Ooijen, R.~Meppelink, J.~M. Vogels and
  P.~van~der Straten, \emph{{Large atom number Bose-Einstein condensate of
  sodium}}, \href{https://doi.org/10.1063/1.2424439}{\emph{Review of Scientific
  Instruments} {\bfseries 78} (2007) 013102}.

\bibitem{Fried1998}
D.~G. Fried, T.~C. Killian, L.~Willmann, D.~Landhuis, S.~C. Moss, D.~Kleppner
  et~al., \emph{{Bose-Einstein Condensation of Atomic Hydrogen}},
  \href{https://doi.org/10.1103/PhysRevLett.81.3811}{\emph{Phys. Rev. Lett.}
  {\bfseries 81} (1998) 3811}.

\bibitem{Greytak2000}
T.~Greytak, D.~Kleppner, D.~Fried, T.~Killian, L.~Willmann, D.~Landhuis et~al.,
  \emph{{Bose–Einstein condensation in atomic hydrogen}},
  \href{https://doi.org/https://doi.org/10.1016/S0921-4526(99)01415-5}{\emph{Physica
  B: Condensed Matter} {\bfseries 280} (2000) 20}.

\bibitem{Panda2022}
C.~D. Panda, M.~Tao, J.~Egelhoff, M.~Ceja, V.~Xu and H.~M\"uller,
  \emph{{Minute-scale gravimetry using a coherent atomic spatial
  superposition}},
  \href{https://doi.org/10.48550/arXiv.2210.07289}{\emph{arXiv:2210.07289}
  (2022) }.

\bibitem{PhysRevA.82.063605}
H.~Pichler, A.~J. Daley and P.~Zoller, \emph{Nonequilibrium dynamics of bosonic
  atoms in optical lattices: Decoherence of many-body states due to spontaneous
  emission}, \href{https://doi.org/10.1103/PhysRevA.82.063605}{\emph{Phys. Rev.
  A} {\bfseries 82} (2010) 063605}.

\bibitem{PhysRevA.84.043628}
G.~Ferrini, D.~Spehner, A.~Minguzzi and F.~W.~J. Hekking, \emph{Effect of phase
  noise on quantum correlations in bose-josephson junctions},
  \href{https://doi.org/10.1103/PhysRevA.84.043628}{\emph{Phys. Rev. A}
  {\bfseries 84} (2011) 043628}.

\bibitem{PhysRevLett.89.140402}
M.~W. Jack, \emph{Decoherence due to three-body loss and its effect on the
  state of a bose-einstein condensate},
  \href{https://doi.org/10.1103/PhysRevLett.89.140402}{\emph{Phys. Rev. Lett.}
  {\bfseries 89} (2002) 140402}.

\bibitem{schelle2009environment}
A.~Schelle, \emph{Environment-induced dynamics in a dilute Bose-Einstein
  condensate}, Ph.D. thesis, Universit{\'e} Pierre et Marie Curie-Paris VI,
  2009.

\bibitem{PhysRevA.60.R29}
S.~Bali, K.~M. O'Hara, M.~E. Gehm, S.~R. Granade and J.~E. Thomas,
  \emph{Quantum-diffractive background gas collisions in atom-trap heating and
  loss}, \href{https://doi.org/10.1103/PhysRevA.60.R29}{\emph{Phys. Rev. A}
  {\bfseries 60} (1999) R29}.

\bibitem{PhysRevA.81.013620}
K.~Paw\l{}owski and K.~Rz\k{a}\ifmmode~\dot{z}\else \.{z}\fi{}ewski,
  \emph{Background atoms and decoherence in optical lattices},
  \href{https://doi.org/10.1103/PhysRevA.81.013620}{\emph{Phys. Rev. A}
  {\bfseries 81} (2010) 013620}.

\bibitem{Brax2011}
P.~Brax, C.~van~de Bruck, A.-C. Davis, B.~Li and D.~J. Shaw, \emph{{Nonlinear
  Structure Formation with the Environmentally Dependent Dilaton}},
  \href{https://doi.org/10.1103/PhysRevD.83.104026}{\emph{Phys. Rev. D}
  {\bfseries 83} (2011) 104026}
  [\href{https://arxiv.org/abs/1102.3692}{{\ttfamily 1102.3692}}].

\end{thebibliography}\endgroup
\bibliographystyle{JHEP}

\end{document}